\documentclass[prl,twocolumn,amsmath,amssymb,graphicx]{revtex4-2}

\usepackage{bm,url,natbib,textcase,color,hyperref,amsmath}

\newcommand{\arXiv}[1]{\href{http://www.arXiv.org/abs/#1}{arXiv:#1}}
\renewcommand{\doi}[2]{\href{http://dx.doi.org/#2}{#1}}
\newcommand{\beq}{\begin{equation}}
\newcommand{\eeq}{\end{equation}}

\DeclareMathOperator{\Tr}{Tr}

\begin{document}

\title{A relation between Krylov and Nielsen complexity}

\author{Ben Craps$^{a}$, Oleg Evnin$^{b,a}$ and Gabriele Pascuzzi$^a$\vspace{2mm}}
\affiliation{$^a$ TENA, Vrije Universiteit Brussel (VUB) and International Solvay Institutes, Brussels, Belgium\vspace{0.5mm}}
\affiliation{$^b$ High Energy Physics Research Unit, Faculty of Science, Chulalongkorn University, Bangkok, Thailand}

\begin{abstract}

Krylov complexity and Nielsen complexity are successful approaches
to quantifying quantum evolution complexity that have been actively
pursued without much contact between the two
lines of research. The two quantities are motivated by quantum chaos
and quantum computation, respectively, while the relevant mathematics
is as different as matrix diagonalization algorithms and geodesic
flows on curved manifolds. We demonstrate that, despite these
differences, there is a relation between the two quantities. Namely,
the time average of Krylov complexity of state evolution can be expressed as a trace
of a certain matrix, which also controls an upper bound on Nielsen complexity
with a specific custom-tailored penalty schedule adapted to the Krylov
basis.

\end{abstract}

\maketitle

Quantifying the complexity of quantum evolution has been a growing topic of research in recent years, driven by two complementary perspectives.
First, it is naturally expected that integrable dynamics 
is intrinsically less complicated than chaotic dynamics, and one may hope that complexity-related measures will provide yet
another insight into the nature of this distinction. Second, any quantum evolution can be seen as tautologically simulating itself,
and this invites the application of computational complexity measures that have emerged from years of research on quantum computing.

The two perspectives we have just described can be referred to as the `quantum chaos' and `quantum computation' perspectives, respectively.
Correspondingly, two different branches of research on quantum evolution complexity are in existence. Krylov complexity, originally introduced in \cite{K},
attempts to quantify how fast operators spread in the space of all possible operators as they evolve. This program is rooted in the quantum chaos
lore and linked to earlier studies of out-of-time-order correlators \cite{OTOC}. Nielsen complexity, on the other hand, emerged in \cite{N1,N2,N3} as a continuum analogue of discrete gate complexity measures in quantum computation. The mathematics involved in defining these two quantities could not be more different. On the Krylov complexity side, the main ingredient is the Lanczos algorithm for matrix tridiagonalization, which creates a useful basis for tracking down the spread of the initial seed operator under the dynamical evolution. On the Nielsen complexity side, the main ingredient is optimization of the length of curves on
the manifold of unitary operators endowed with an anisotropic metric (this metric captures the relative difficulty of performing some unitary transformations on a physical system, for example those involving changing the state of many particles at once).

Perhaps not surprisingly for two quantities so different in their origins and in the relevant mathematics involved, research on Krylov complexity \cite{Kstate, Kint1,Kint2,Kdep,plateau,plateau1,plateau2} and Nielsen complexity \cite{evolcompl1,evolcompl2,bound,complint} of quantum evolution has developed in parallel \cite{discl}, with very little contact beyond descriptive qualitative comparisons of the outcomes. Our purpose in this Letter is to spell out a mathematical framework that unites these two quantities.

Throughout, we shall consider a quantum system with Hamiltonian $H$, and Hilbert space of finite dimension $D$ \cite{finite}. The evolution operator will thus be a $D\times D$ unitary matrix residing in the group manifold $SU(D)$. We furthermore define the energy eigenvalues $E_n$ and eigenstates $|n\rangle$ for future use:
\beq
    H |n\rangle = E_n |n\rangle \hspace{1cm} n=0, \ldots ,D-1.
\eeq

{\it Krylov complexity and its average.}---
The original definition of Krylov complexity in \cite{K} tracked the Heisenberg evolution of quantum operators. We shall be focusing here on its closely related analogue introduced in \cite{Kstate} that applies the same protocol to the Schr\"odinger evolution of quantum states. (Terms like `Krylov complexity of states' or `spread complexity' may be used.)

One starts with an initial vector $|v_0\rangle$ and lets it evolve as $e^{-iHt}|v_0\rangle$. The qualitative question is: how many extra vectors does one need to effectively capture the evolution as time goes on, and how rapidly does this number increase with time? (Evidently at $t=0$, $|v_0\rangle$ would suffice by itself, while at late times one would likely need a complete basis.)

To give these questions a concrete expression, one introduces 
the Krylov basis $|v_j\rangle$, generated from the Hamiltonian $H$ and the initial state $|v_0\rangle$ via the Lanczos algorithm:
\beq
    |w_{j+1}\rangle = (H-a_j)|v_j\rangle - b_j |v_{j-1}\rangle, \hspace{0.4cm} |v_j\rangle = \frac{1}{b_j} |w_j\rangle.
    \label{Laczstep}
\eeq
Here, the Lanczos coefficients $a_j$ and $b_j$ are defined by
\beq\label{abcoeff}
    a_j = \langle v_j |H| v_j \rangle, \hspace{0.4cm} b_j = \sqrt{\langle w_j | w_j \rangle}, \hspace{0.4cm} b_0 = 0.
\eeq
The basis is constructed to be orthonormal. In fact, the Lanczos algorithm is nothing but Gram-Schmidt orthonormalization applied to the Krylov sequence $H^j|v_0\rangle$. Generically, the Krylov basis spans the full space of dimension $D$, so the algorithm terminates after $D-1$ steps.

We can describe the time evolution of $|v_0\rangle$ in the Krylov basis:
\beq
    |\phi (t)\rangle = e^{-iHt}|v_0\rangle = \sum_{j=0}^{D-1} \phi_j (t) |v_j \rangle.\label{phidecomp}
    \eeq
The Hamiltonian is tridiagonal in the Krylov basis, as can be deduced from (\ref{Laczstep}). Thus, given a Hamiltonian $H$ and an initial `seed' state $|v_0\rangle$, the evolution is recast into a 1d 
nearest-neighbor hopping model, with $\phi_j$ being the value of the wavefunction at site $j$. (Early appeals
to using such tridiagonal representations for physical Hamiltonians can be seen in
\cite{hist1,hist2}.)
    
Krylov complexity is then designed as a measure of the average position of the hopper along the chain at time $t$. (At $t=0$, it is evidently localized at site 0.) Specifically, with a sequence of positive nondecreasing weights $w_j$, we define
\beq
    C_K (t) = \sum_{j=0}^{D-1} w_j |\phi_j (t) |^2.
\eeq
In practical applications, one often chooses $w_j=j$, so that $C_K$ is literally the average position. If this value does not grow much, one expects that $|\phi (t)\rangle$ in (\ref{phidecomp}) is well-approximated by the first few terms in the sum, making it `simple.'

Krylov complexity typically grows at early times, eventually reaching a plateau. This plateau has been tested as a valid indicator of integrable vs.\ chaotic properties of the underlying system \cite{Kint1,Kint2}, though the procedure shows sensitivity to the seed of the Lanczos algorithm \cite{Kdep, plateau1}. A good way to estimate the plateau height is to compute, following \cite{Kint1,Kint2}, the all-time average of $C_K$. For that, we write
\beq
    \phi_j(t) = \sum_{n=0}^{D-1} e^{-iE_n t}\langle v_j |n\rangle \langle n|v_0\rangle,
\eeq
and hence
$$
    |\phi_j(t)|^2 = \sum_{n,m=0}^{D-1} e^{-i(E_n - E_m) t} \langle v_j | n \rangle \langle n | v_0 \rangle \langle v_0 | m \rangle \langle m | v_j \rangle.
$$
Then, for a generic spectrum with nondegenerate eigenvalues,
\beq
    \overline{|\phi_j|^2} \equiv \lim_{T \to \infty} \frac{1}{T} \int_{0}^{T} |\phi_j(t)|^2 dt
= \sum_{n=0}^{D-1} |\langle v_j | n \rangle|^2\, |\langle n | v_0 \rangle|^2,
\eeq
and thus the time average of Krylov complexity is expressed as
\beq\label{CKav}
\begin{split}
    \overline{C_K} = \lim_{T \to \infty} \frac{1}{T} \int_{0}^{T} C_K(t) dt = \sum_{j=0}^{D-1} w_j \,\overline{|\phi_j|^2} \\ 
    = \sum_{n,j=0}^{D-1} w_j\,|\langle v_j | n \rangle|^2\, |\langle n | v_0 \rangle|^2.\hspace{5mm}
\end{split}
\eeq
This is the main quantity we shall work with when building connections with the Nielsen complexity formalism \cite{lnczs}.\vspace{1mm}

{\it Nielsen complexity and its bound.}---
Defining Nielsen complexity starts with picturing any curve $U(\tau)$ on the manifold $SU(D)$ of $D\times D$ unitary matrices as a `program.'
The complexity of this program is then the length of the curve, and the complexity of a given target unitary $U$ is the minimum of this length over all curves connecting $U$ to the identity matrix. This is a natural continuum analogue of the standard discrete gate complexity in quantum computation, with the continuum length replacing the counting of the number of discrete gate operations. There is a controllable relation between Nielsen and gate complexity for qubit systems \cite{N1,N2,N3}.

Which metric should we use on the manifold of unitaries to compute the length? The most naive guess would be the isotropic bi-invariant metric that treats all directions equally. To define it mathematically,
introduce the Hermitian velocity of the curve $U(\tau)$ as
\beq
    V(\tau) = i \,\frac{dU}{d\tau} U^\dag.
\eeq
The bi-invariant length of the curve segment between $\tau$ and $\tau+d\tau$ is then simply
\beq
    ds^2_{\text{bi-inv}}=\Tr(V^\dag V)d\tau^2.
\eeq
All geodesics of the bi-invariant metric have constant velocities, and therefore the minimization problem involved in the definition of Nielsen complexity can be solved exactly. The results are (unsurprisingly) disappointing, however, since this computation assigns the same value of complexity to all physical systems with the same Hilbert space dimension \cite{bound}. The isotropic bi-invariant metric is simply too naive, as it fails to recognize that some unitary transformations are more difficult to implement than others. This ingredient is essential for complexity definitions in the context of quantum computation, just as restricting the set of allowed gates is essential for defining gate complexity.

One must then look for more sophisticated assignments of the metric on the manifold of unitaries that will bring out the distinction between different kinds of dynamics. To this end,
we introduce a complete basis \cite{nonherm} of generators $T_\alpha$, orthonormal with respect to the inner product $\Tr(T^\dag_\alpha T_\beta)$. We can then decompose the velocity as
\beq
    V = V^\alpha T_\alpha, \hspace{1cm} V^\alpha = \Tr(T^\dag_\alpha V).
\eeq
The Nielsen complexity metric introduces different penalty factors $\mu_a$ for the various directions $T_\alpha$:
\beq
    ds^2 = d\tau^2\, \sum_{\alpha} \mu_\alpha \left|\Tr(T_\alpha^\dag V)\right|^2.\label{complmetr}
\eeq
The bigger the penalty factor $\mu_a$, the more difficult it is to move in that particular direction. The penalty factors are typically chosen on the basis of some locality properties (for example, acting on only a certain number of adjacent spatial sites, or acting only on a given number of particles at once). A common choice is to introduce a threshold so that all operators above the threshold are `hard' (with the same large $\mu_a$) and all those below the threshold are `easy' (with $\mu_a$ set to 1). For the purposes of making contacts with Krylov complexity, we shall keep the penalty factors $\mu_a$ completely general.

While the inclusion of penalty factors gives the metric (\ref{complmetr}) a chance to distinguish different types of dynamics, it also renders the minimization problem involved into the definition of Nielsen complexity largely intractable. The state of the art is that the geodesic equation has been solved for such metrics for the case of three qubits \cite{N3}. What is even more complicated is finding actual geodesics connecting two prescribed points \cite{4qbit}, and minimizing the length over all such geodesics. More importantly, in cases of physical interest, the Hilbert space dimensions is orders of magnitude higher than for the three-qubit system, and all the difficulties multiply at a crippling rate as the number of dimensions increases, making direct evaluation of Nielsen complexity impossible.

A practical strategy to deal with these issues has been put forth in \cite{bound,complint}. If it is out-of-reach to minimize the distance over all possible curves using the metric (\ref{complmetr}), we can settle on computing an upper bound on Nielsen complexity by minimizing the distance over a prescribed infinite family of curves.

A simple and effective family of curves for this purpose \cite{bound} is constant velocity curves $e^{-iV\tau}$. While all of these start at the identity at $\tau=0$, to ensure that they connect to the desired evolution operator $U=e^{-i H t}$ at $\tau=t$, that is $e^{-i V t}=e^{-i H t}$, we must impose 
\beq\label{Vdef}
    V = \sum_n \left( E_n -\frac{2 \pi k_n}{t}\right) |n\rangle \langle n | ,
\eeq
where $k_n$ are $D$ independent integers. (For simplicity, we are assuming a nondegenerate energy spectrum, as expected in generic systems. See \cite{complint} for how to treat degenerate spectra.) From (\ref{complmetr}) and (\ref{Vdef}), a bound on Nielsen complexity $C_b$ is then given by a minimization over the $D$-dimensional hypercubic lattice $k_n$ as
\beq\label{Cb}
    C_{b}(t)=2\pi \min_{\vec{k} \in \mathbb{Z}^D} \sqrt{ (\vec{y}, \mathbf{Q}\vec{y})},\qquad \vec{y}\equiv \frac{\vec E t}{2 \pi} 
    -\vec{k},
\eeq
with $\vec{E}\equiv (E_0,E_1,\ldots E_{D-1})$, $\vec{k}\equiv (k_0,k_1,\ldots,k_{D-1})$ and the matrix \cite{Qdiscl}
\beq\label{Qdef}
Q_{nm}\equiv \sum_\alpha
\mu_\alpha
 \langle n| T_\alpha|n\rangle \langle m|T_\alpha^\dag |m\rangle.
\eeq

The minimization problem (\ref{Cb}) has a natural geometric interpretation. Up to constant factors, one is simply asking about the distance from the point $\vec{E}t/2\pi$ to the nearest point of the integer hypercubic lattice $\mathbb{Z}^D$, with the distances measured using not the standard Euclidean metric, but rather the (position-independent) `skewed' metric (\ref{Qdef}). This is known as the {\it closest vector problem}, and it has been discussed extensively in the mathematics literature, specifically in relation to lattice-based cryptography \cite{lattice}. Effective algorithms exist for finding approximate solutions to this problem, making the minimization problem (\ref{Cb}) computationally tractable, unlike the original definition of Nielsen complexity. All of this has been implemented in practice and applied to a broad range of physical Hamiltonians in \cite{bound,complint}, which can be consulted for technical details.  While our reliance on constant velocity curves as a proxy for minimization over all curves may seem rather {\it ad hoc} at first sight, it has been tested in practice and is able to produce meaningful results \cite{bound, complint}. In particular, the upper bound (\ref{Cb}) consistently assigns lower values to complexities of integrable Hamiltonians than to chaotic ones. Furthermore, a direct relation between local (few-body) conservation laws (a hallmark of integrability) and the properties of the $Q$-matrix (\ref{Qdef}) has been manifested in \cite{complint}.

Just as for Krylov complexity, some generic features are expected for the time dependence of Nielsen complexity: initial growth and late-time saturation. And as for Krylov complexity, the height of this late-time plateau is of particular interest. In the case of the bound (\ref{Cb}), it can be understood from the following heuristic argument. In high numbers of dimensions, distances from a randomly chosen point to a lattice tend to take essentially deterministic values, as one can see by elementary means for Euclidean distances from hypercubic lattices; a practical summary can be found in \cite{bound}. As the vector $\vec{E}t/2\pi$ grows and reaches far outside the unit cell where it started, it is likely to behave as a generic point in space with respect to its distance from the lattice in (\ref{Cb}). Hence, its actual distance from the lattice (which is, by definition, the plateau height) will closely track the average distance \cite{average}. This leads to the following estimate of the late-time plateau value $C_p$ of (\ref{Cb}):
\beq\label{Cp}
C_p=2\pi\int_{0\le x_n\le 1} d\vec{x}\min_{\vec{k} \in \mathbb{Z}^D} \sqrt{ (\vec{x}-\vec{k}, \mathbf{Q}(\vec{x}-\vec{k}))}.
\eeq
This estimate is completely controlled by the $Q$-matrix (\ref{Qdef}), which will continue to play a central role in our story.

{\it Krylov complexity $\leftrightarrow$ Nielsen complexity.}--- With the above setup, we are in a position to make our key observations.
The starting points for formulating Krylov and Nielsen complexity are very different (Hamiltonian tridiagonalization by linear transformations, and geodesic optimization on the curved manifold of unitaries endowed with an anisotropic metric), so that it is not even immediately clear how to attempt mapping the two structures into each other. At the same time, once we focus on the all-time average of Krylov complexity (\ref{CKav}) and the late-time value for the upper bound on Nielsen complexity given by (\ref{Cb}), the situation no longer appears hopeless. Indeed, the expressions (\ref{CKav}) and (\ref{Qdef}) are suggestively similar, being both quartic in the Hamiltonian eigenvector components. To complete the picture, it remains to spell out an explicit relation between the two sets of formulas. 

To do so, we note that (\ref{CKav}) can be recast in the form
\beq
\overline{C_K} = \Tr \mathbf{q},
\eeq
with
\beq\label{qdf}
    q_{nm}= \sum_{j=0}^{D-1} \frac{w_j}{2}\Big(\langle n| v_0 \rangle \langle v_j | n \rangle \langle m| v_j \rangle \langle v_0 |m\rangle + \mbox{c.c.}\Big).
\eeq
Can we understand this $q$-matrix as a special case of the general $Q$-matrix (\ref{Qdef}), which would immediately connect us to Nielsen complexity? The answer is yes, and one simply needs to provide an identification of the penalty factors $\mu_a$ and the operator basis $T_a$ that reduces (\ref{Qdef}) to (\ref{qdf}).

The relevant assignments can be summarized in the following table:
\beq\label{penalty}
\begin{tabular}{ |c|c| } 
 \hline
 $T_a$ & \hspace{2mm}$\mu_a$ assignment\hspace{2mm}  \\ 
  \hline
$|v_0\rangle\langle v_0 |$& $w_0$ \\ 
  \hline\hspace{0.5mm}
$|v_0\rangle\langle v_j |$ or $|v_j\rangle\langle v_0|$ with $j\ge 1$\hspace{0.5mm} & $w_j/2$  \\ 
  \hline
$|v_i\rangle\langle v_j |$ with $i,j\ge 1$  & 0 \\ 
 \hline
\end{tabular}
\eeq
One can check that, with these assignments, the general $Q$-matrix given by (\ref{Qdef}) becomes identical to (\ref{qdf}). A link between Krylov and Nielsen complexity has thus been established: {\it The all-time average of Krylov complexity is the trace of the  $q$-matrix (\ref{qdf}). An upper bound on the plateau value of Nielsen complexity with the penalty schedule (\ref{penalty}) is given by the average distance from an integer lattice (\ref{Cp}) in a space where the metric $\mathbf{Q}$ is set equal to the same matrix $\mathbf{q}$.} This relation is not an equality of the two quantities \cite{triangle}, but rather a way to express them as explicit functions of the same $q$-matrix. The expressions are similar in spirit, however: for example, if the eigenvalues of $\mathbf{q}$ grow, its trace evidently increases, but so does the average distance estimate, since the distance growth in different directions is controlled by the eigenvalues.

The penalty schedule (\ref{penalty}) is rather peculiar in that it assigns zero penalty to all generators not involving the Krylov seed vector $|v_0\rangle$. This penalty schedule nonetheless passes an important sanity check: while the commutators of the operators in the second and third lines of (\ref{penalty}) generate a full basis of operators (this property is called `bracket generating' in the language of sub-Riemannian geometry \cite{subriemannian} that is often evoked in the context of Nielsen complexity \cite{howsmooth}), nested commutators of the zero-penalty operators in the last line of (\ref{penalty}) will never produce anything involving $|v_0\rangle$, and hence do not provide a complete basis. For that reason, while moving in the zero-penalty directions does not result in any length increase, it is impossible to reach the target while moving in those directions alone, and thus all the minimization problems that define Nielsen complexity remain meaningful. 

It is hardly surprising that a rather peculiar penalty schedule had to be used to make Nielsen complexity capture the behavior of Krylov complexity. Indeed, Krylov complexity only tracks the evolution of a single seed vector $|v_0\rangle$, while Nielsen complexity is sensitive to the entire evolution operator. A sort of blinding device had to be applied to the latter to make it mimic the former. It is natural to think of the zero penalties in (\ref{penalty}) as such a blinding device.

Note furthermore that the assignment of the growing sequence $w_j$ as penalties for the operators $|v_0\rangle\langle v_j |$ is very natural. Indeed, in a typical construction of Krylov complexity, the initial seed will be very simple (for example, a state where only one spin of a spin chain is excited). A typical Hamiltonian, on the other hand, also has a simple structure of interactions (for example, a sum of terms each of which couples only two spins). Such constructions are prevalent in \cite{K,Kint1,Kint2,Kdep} and other literature on the subject. In this situation, higher $|v_j\rangle$ will be progressively more and more complicated states (say, with more and more spins excited). Correspondingly, it is natural to think of the operators $|v_0\rangle\langle v_j |$ as becoming more complicated with growing $j$, assigning them an increasing sequence of penalties. (This naive intuition does not work, however, for the operators $|v_i\rangle\langle v_j |$, which must be assigned zero penalties to make the construction work, as already remarked in the previous paragraph.)

The framework we have described provides a stepping stone to explore a range of similar relations for analogous quantities. A natural question is whether the Krylov complexity for operators, as originally defined in \cite{K}, can be given a similar treatment.
Another question is whether Nielsen complexity of states \cite{state1,state2}, defined as the minimum of the Nielsen complexity for unitaries over all unitaries that convert a given reference state to the desired state, can meaningfully enter the game, and in particular provide additional context for the penalty schedule (\ref{penalty}).
It could also be interesting to consider further averaging of (\ref{CKav}) over a suitably chosen set of initial vectors, and explore how that affects (\ref{penalty}), possibly producing more conventional penalty schedules.

To sum up, we have displayed a relation between Krylov complexity of states and an upper bound on the Nielsen complexity of the evolution operator. Both quantities end up being expressed through a specific matrix defined by (\ref{qdf}).
Krylov complexity and Nielsen complexity take rather different inputs for their definitions, and we had to spell out explicitly how these inputs are to be matched so as to make our construction work. Krylov complexity requires specifying an initial seed vector $|v_0\rangle$, as well as weights $w_j$ for the contributions of higher Krylov components of the wavefunction. Nielsen complexity requires specifying a basis of generators on the manifold of unitaries and assigning to them penalty factors. Within the correspondence we established, the Krylov seed and its Krylov basis generated using the Lanczos algorithm are converted into the generator basis on the Nielsen complexity side, while the weights $w_j$ are converted into penalty factors.

Relating Krylov and Nielsen complexity connects two very different pictures of quantum evolution: the extent of spread of states over the Hilbert space with the flow of time on the one side, and quantum algorithms and quantum simulations on the other side. It has in particular been proposed (see, for instance, the conclusions of \cite{alg}) that minimization problems that define Nielsen complexity can practically contribute to finding optimal quantum simulation algorithms. A relation between such algorithms and quantities that control evolutionary spread of states is thus another implication of the findings we report here.

\vspace{2mm}
%%%%%%%%%%%%%%%%%%%%%%%%%%%%%%%%%
\noindent{\bf Acknowledgments:}
We thank Vijay Balasubramanian, Marine De Clerck and Andrew Rolph for discussions on related subjects.
This work has been supported by the Research Foundation Flanders (FWO) through project G012222N, and by Vrije Universiteit Brussel through the Strategic Research Program High-Energy Physics.  OE is supported by Thailand NSRF via PMU-B (grant number B13F670063). GP is supported by a PhD fellowship from FWO.


\begin{thebibliography}{99}
\bibitem{discl}Due to the vastness on the bibliography on these topics, we only cite a selection of papers, chosen because they play a direct role in our construction.
\bibitem{lnczs}It may appear paradoxical at first sight that (\ref{CKav}), which is a key object of our study, does not depend explicitly on the Lanczos coefficients (\ref{abcoeff}), but only on the Lanczos basis $|v_j\rangle$. Much of the discussion of the phenomenology of Krylov complexity in the literature revolves around the properties of Lanczos coefficients. Note, however, that the diagonalization of the Hamiltonian in the Lanczos basis depends explicitly on the values of the Lanczos coefficients, and therefore the Lanczos coefficients are encoded in the Lanczos basis components of the Hamiltonian eigenstates $\langle v_j|n\rangle$ appearing in (\ref{CKav}). This resolves the apparent paradox.
\bibitem{nonherm}It is convenient in our present setting not to assume that the generators are Hermitian. This, of course, does not bring in any essential differences, since any non-Hermitian generator, together with its Hermitian conjugate, can be converted into a pair of Hermitian generators. As a result, some formulas in this Letter have additional Hermitian conjugation inserted in comparison with the presentation in \cite{bound,complint}.
\end{thebibliography}
\end{document}